\RequirePackage{fix-cm}

\documentclass[smallextended]{svjour3}       

\smartqed  
\usepackage{graphicx}
\usepackage{adjustbox}
\usepackage{amssymb}
\usepackage{float}
\floatstyle{plaintop}
\restylefloat{table}
 \usepackage{caption}
\usepackage[ruled,vlined]{algorithm2e}
\usepackage{float}
\restylefloat{table}
\usepackage[fleqn]{amsmath}
\usepackage{amssymb}

\begin{document}

\title{Iterated Greedy Algorithms for the Hop-Constrained Steiner Tree Problem
}


\author{Farzane Yahyanejad         \and
        Bahram Sadeghi Bigham 
}


\institute{Farzane Yahyanejad \at
	      Department of Computer Science, University of Illinois at Chicago, IL 60607, USA
              f.yahya2@uic.edu \\
              Tel.: +412-951-4151\\
              \email{f.yahya2@uic.edu}           
           \and
           Bahram Sadeghi Bigham \at
              Department of Computer Science, Institute for Advance Studies in Basic Sciences, Zanjan, Iran
              \email{b$\_$sadeghi$\_$b@iasbs.ac.ir}
}

\date{Received: date / Accepted: date}

\maketitle

\begin{abstract}
The Hop-Constrained Steiner Tree problem (HCST) is challenging NP-hard problem arising in the design of centralized telecommunication networks where the reliability constraints matter. In this paper three iterative greedy algorithms are described to find efficient optimized solution to solve HCST on both sparse and dense graphs. In the third algorithm, we adopt the idea of Kruskal algorithm for the HCST problem to reach a better solution. This is the first time such algorithm is utilized in a problem with hop-constrained condition. Computational results on a number of problem instances are derived from well-known benchmark instances of Steiner problem in graphs. We compare three algorithms with a previously known method (Voss's algorithm \cite{14}) in term of effectiveness, and show that the cost of the third proposed method has been noticeably improved significantly, 34.60$\%$ in hop 10 on dense graphs and 3.34$\%$ in hop 3 on sparse graphs.
\end{abstract}

\section{Introduction}
\label{intro}
Consider an undirected graph G=(V, E) with $v=\{1, 2,... , n\}$, where specific vertex 1 is root vertex and edge set is $E=\{(i,j): i,j\in V,i<j\}$, that each edge has an associated cost $c_{ij}\geq 0$. Also, we define a positive integer $H$ to show the maximum allowance distance of edge from the root. Let $T$ be a tree in $G$, and let $V(T)$ be the set of vertices belonging to $T$. Our goal is finding a tree $T$, rooted at vertex 1, subject to constraints and the number of edges $h_i$ between $i\in V (T)$ and the root vertex limited to the maximum value H. 

Since HCST is a generalization of Steiner tree problem, it is Np-hard problem\cite{109}. We consider hop constraints in graph because of reasons such as reliability or transmission delay in networks. The extensive uses of hop constraints in various settings have been proposed in the literature \cite{103,102,101,107,108}.
Intensive researches on the Minimum Spanning Tree problem with hop constraints (HCMST), which is a special case of the HCST problem where all vertices in the graph are terminals exist.
 Many surveys regarding of HCMST problem can be found in \cite{5,111,110}. Though, there aren't much attention to Steiner tree problem with hop constraints. In \cite{7}, Gouveia has mentioned HCST problem with developing a strengthened version of a multi-commodity flow model for the minimum spanning tree problem. 
 The LP lower bounds of this model are equal to the ones from a Lagrangian relaxation approach of a weaker MIP model introduced by Gouveia in \cite{8}. Voss presents MIP formulations based on Miller-Tucker-Zemlin subtour elimination constraints \cite{14}. The formulation is then strengthened by disaggregation of variables indicating used edges. 
 The author develops a simple heuristic to find starting solutions and improves them with an exchange procedure based on tabu search. Gouveia Also gives a survey of hop-indexed tree and flow formulations for the hop constrained spanning and Steiner tree problem in \cite{9}. 
 Costa et. al. give a comparison of three methods for a generalization of the HCSTP which is called Steiner tree problem with revenues, budget and hop constraints (STPRBH) in \cite{1}. The considered methods comprise greedy algorithm, destroy-and-repair method and tabu search approaches. 
 Computational results are reported for instances with up to 500 vertices and 12500 edges. Costa et al. in \cite{2} introduce two new models for he STPRBH. Both models are based on the generalized sub-tour elimination constraints and a set of exponential size hop constraints. The authors provide a theoretical and computational comparison with two models based on Miller-Tucker-Zemlin constraints presented in Gouveia \cite{10} and Voss \cite{14}.  Theoretical and computational comparisons of flow-based vs. path-based mixed integer programming models for HCST problem are presented by Gouveia et al.. They propose formulations to solve the problem with promising optimality and implement branch-and-price algorithms for all of the formulations\cite{11}.
Boeck et al. used layered graphs for hop constrained problems to build extended formulations by techniques presented to reduce the size of the layered graphs\cite{65}. They also presented variation of this problem arising in the context of multicast transmission in telecommunications.
Dokeroglu et al. recently proposed novel self-adaptive and stagnation-aware breakout local search algorithm for the solution of Steiner tree problem with revenue, budget and hop constraints with parallel algorithms \cite{15}.
In this paper, we focus on efficient optimization iterative greedy algorithms to find efficient solution for HCST. The new algorithms find a feasible solution for HCST are based on Voss's approach in \cite{14}. The main feature of our last proposed algorithm is that it uses the idea of Kruskal algorithm in the problem with hop constraint for the first time and enhances the results of this NP-hard problem in polynomial time.

The rest of this paper is organized as following. In Section 2 we formulate a model for HCST problem. Section 3 and 4 provide two complement greedy algorithms to solve HCST. In Section 5 we propose Non Root Based Insertion Algorithm based on the idea of Kruskal algorithm\cite{3}. In Section 6 we report extensive set of computational experiments with our iterative greedy algorithms on well known benchmark, consisting variant kinds of graphs. Finally, Section 7 presents concluding remarks .

\section{Mathematical formulation}
\label{MathModel}
A non-directed graph $G = (V, E)$ is given with edge cost $c_{ij}\geq 0$, $(i, j)\in E$. Let consider the set $Q\subseteq V$ represent basic vertices. To obtain a subgraph with minimum cost, other vertices could be involved. Such vertices are called Steiner vertices $(S=V-Q)$. Consider $x^p_{ij}$ as an edge $(i, j) \in E$ that shows the $p^{th}$ position from the root and $\delta (S)$ as a set of edges with one vertex from $S$. The HCST model can be written as follow:
\begin{equation}
min \sum_{p=1}^{H}\sum_{(i,j)\in E}c_{ij}x_{ij}^{p} 
\end{equation}
\begin{equation}
\sum_{p=1}^{H}\sum_{(i,j)\in \delta (S)}x_{ij}^p\geq 1 \hspace{1cm}\forall S \subseteq V, S\cap Q\neq \emptyset , (V\backslash S)\cap Q\neq \emptyset
\end{equation}
\begin{equation}
x_{ij}^p\in \{0, 1\} \hspace{1cm}\forall (i, j)\in E
\end{equation}
The goal is to find a subgraph with minimum cost due to the hop constraints; Constraints in (2) present the connectivity constraints of basic vertices attaching to at least one edge of the vertices from $S$ and index $p$ is used to avoid creation of any cycle.
\section{Minimum-Hop Iterative Greedy (MinHIG) Algorithm}
\label{}
In this section, we present an iterative greedy algorithm for HCST problem. The basic idea is from \cite{14}. The first three steps of the algorithm use the generalization of prime algorithm \cite{12} considering the path with number of hop instead of a direct edge between two vertices (Algorithm 1). 

We start with a partial solution $G=({root},\emptyset)$ that just contains the root. As we mentioned before, $Q$ is the set of all basic vertices. At each step the set $T$ is equal to $Q\backslash G$ and $H$ is the maximum allowed number of hops that the root can connects other vertices in the tree.
Suppose $V_G$ denote the vertex set of $G$ and $V_T$ denote the vertex set of $T$.  Cost of path $p(u,v)$ between vertices $u$ and $v$ is presented by $d_{uv}$. Also, for every vertex $v$, we define $U_v$ equal to the number of needed hops to reach $v$ from the root. For initialization, we set $U_{root}$ to zero. Phase 1 in MinHIG algorithm finds $U_v$ for all vertices $v$.

\begin{algorithm}[H]
\SetKwInOut{Input}{input}\SetKwInOut{Output}{output}
\Input{An undirected graph.}
\Output{Find U values.}
{\bf Step 1}. {\bf Initialization} $G=(\{root\},\emptyset)$;

\While{$Q\not\subseteq G$}{
{\bf Step 2}. {\bf Find} vertices $u^*\in V_G$, $v^*\in V_T$ and path $P(u^*,v^*)$, where 

$U_v^* = U_u^* +|P(u^*,v^*)|\leq H$,

$d_{u^* v^*}=min⁡\{d_{uv} |u\leq V_G ,v^*\in V_T\}$;

{\bf Step 3}. {\bf Add} the vertices and edges of path $P(u^*,v^*)$ to G,

{\bf update} U values of the vertices of path $P(u^*,v^*)$.
}
\caption{MinHIG algorithm (Phase 1)}
\end{algorithm}

\begin{algorithm}[H]
\SetKwInOut{Input}{input}\SetKwInOut{Output}{output}
\Input{U values.}
\Output{An efficient feasible solution.}
{\bf Step 1}. {\bf Initialization} $Tree=(\{root\},\emptyset)$, h=1;

\While{$h\neq H$}{

{\bf Step 2}. {\bf Find} vertices $u^*\in Q$, $v^*\in V_{Tree}$, and path $P(u^*,v^*)$, where 

$U_u^*=h$,

$d_{u^* v^*}=min⁡\{d_{u^*v} |v\in V_{Tree}, U_{u^*v^*} = U_u^* +|P(u^*,v^*)|\leq H\}$;

{\bf Step 3}. {\bf Add} the vertices and edges of path $P(u^*,v^*)$ to Tree.

$h\leftarrow h+1$
}
\caption{MinHIG algorithm (Phase 2)}
\end{algorithm}
Note that in MinHIG algorithm, we suppose that in the efficient optimal solution, vertices with high $U$ values are connected to vertices with low $U$ values.

In phase 2 in MinHIG, first we add the root vertex to the final tree $Tree$.
Then, we start with $h=1$ and and add it until reach maximum hop constraint $h=H$. Vertices with fewer number of hops have been added to the tree earlier. Therefor, with constraint $U_u^* +|P(u^*,v^*)|\leq H$, we find $P(u^*,v^*)$, which is the shortest path among all paths of vertex $u^*$ to the all previous added vertices to $Tree$. Thus, all edges and vertices of this path will be added to the $Tree$. We repeat this procedure until $Tree$ contains all basic vertices. 
Figure 1 shows an example of instances where MinHIG gets a solution while Voss's algorithm \cite{14} doesn't work.
\begin{figure}[h]
\centering
\includegraphics[scale=0.3]{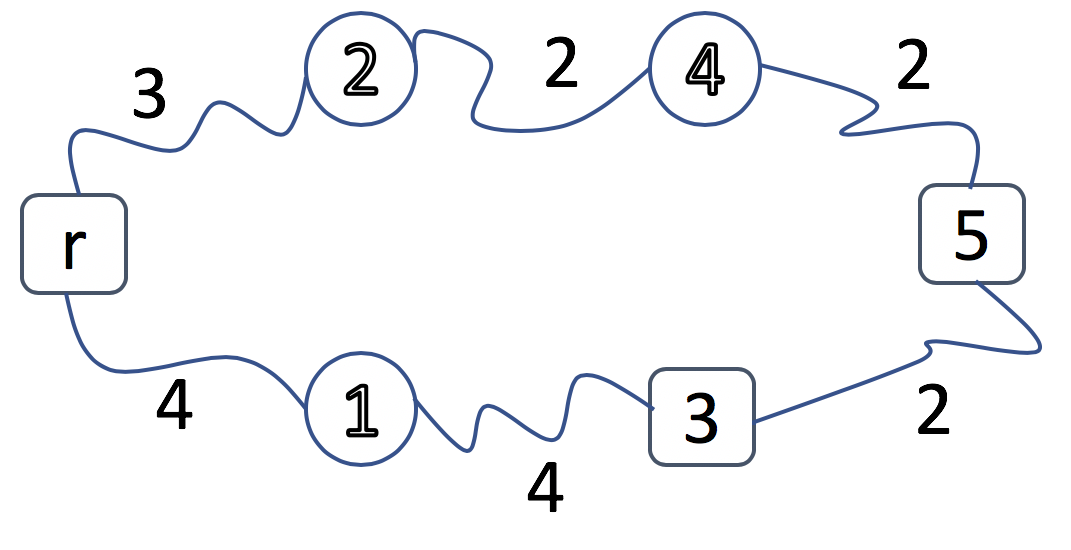}
\caption{An illustrative example to build HCST with 3 mandatory vertices and 3 Steiner vertices(Circles).}
\label{figNNfdsds}
\end{figure}
For Figure 1 consider maximum allowed $H$ is 3. With Voss's algorithm \cite{14}, at first, path $P(r\rightarrow 2\rightarrow 4\rightarrow 5)$ with cost 7 is added to the tree. Then path $P(r\rightarrow 1\rightarrow 3)$ with cost 8 will be added. The final Steiner tree is built has cost 15,  although, it is obvious that the optimal cost of the Steiner tree is 10. In Step 3 in phase 2 of Algorithm 1 we have tree reconstruction which gives us the cost of 10 for this example. 
Suppose that vertices with higher amount of $U$ should be connected to the vertices with lower $U$ values. Therefore, in Step 3 in phase 2, without considering set G, vertices with less U values are added to the final tree. In fact, when we add each vertex, we are sure that other vertices with lower $U$ have been added to the tree earlier and if the current vertex should be connected to one of the basic vertices, we are sure that all possible vertices have been already added to the tree. 
With implementation of MinHIG on this specific example, after first phase values of $U_r=0, U_1=1, U_3= 2, U_2 =1, U_4= 2,$ and $U_5 =3$ have been obtained. After the root is added, basic vertex 3 with the lowest $U$ value among basic vertices with shortest path $P(r\rightarrow 1\rightarrow 3)$ and cost 8 will be added to the tree. Then, the next basic vertex which has not been added to the $Tree$ yet and has minimum value $U$ is vertex 5 with shortest path $P(3\rightarrow 5)$ and cost 2. At the end, the cost of $Tree$ would be 10.
\section{Maximum Hop Iterative Greedy (MaxHIG) Algorithm}
\label{}
In MinHIG algorithm, it was assumed that in the Steiner tree, the vertices with higher $U$ are connected to vertices with lower $U$ ,but this approach doesn't cover all groups of graphs. There are other kinds of graphs in which the vertices by lower hops are connected to the path of the other vertices with more hops.
For example, consider Figure 2 where maximum allowed number of $H$ is 4.
\begin{figure}[h]
\centering
\includegraphics[scale=0.3]{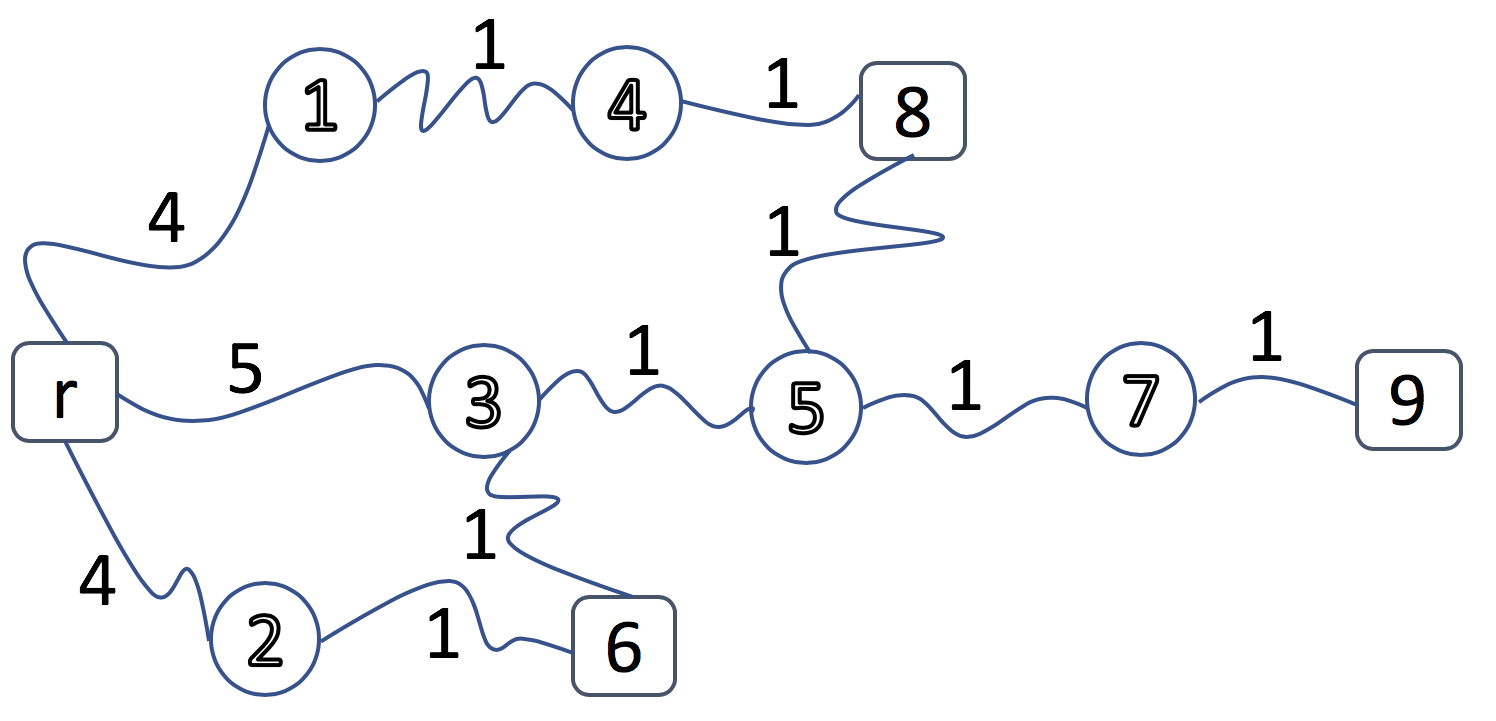}
\caption{An illustrative example to build HCST with 4 mandatory vertices and 6 Steiner vertices}
\label{figNNfdsds}
\end{figure}
By MinHIG algorithm, first we add path $P(r\rightarrow 2\rightarrow 6)$ with cost 5 to the $Tree$. Then, path $P(r\rightarrow 1\rightarrow 4\rightarrow 8)$ with cost 6 and path $P(r\rightarrow 3\rightarrow 5\rightarrow 7\rightarrow 9)$ with cost 8 will be added to the $Tree$ respectively. Therefor, the result cost of the achieved Steiner tree by this algorithm would be 19, while the optimal cost is 10. 

\bigskip
\begin{algorithm}[H]
\SetKwInOut{Input}{input}\SetKwInOut{Output}{output}
\Input{An undirected graph.}
\Output{An efficient feasible solution.}
{\bf Run Algorithm 1}

{\bf Step 1}. {\bf Initialization} $Tree=(\{root\},\emptyset)$, h=H;

\While{$h\neq 1$}{

{\bf Step 2}. {\bf Find} vertices $u^*\in Q$, $v^*\in V_{Tree}$, and path $P(u^*,v^*)$, where 

$U_u^*=h$,

$d_{u^* v^*}=min⁡\{d_{u^*v} |v\in V_{Tree}, U_{u^*v^*} = U_u^* +|P(u^*,v^*)|\leq H\}$.

{\bf Step 3}. {\bf Add} the vertices and edges of path $P(u^*,v^*)$ to Tree.
}
\caption{MaxHIG algorithm}
\end{algorithm}
\bigskip
For these types of graphs we substitute algorithm MinHIG phase 2 (Algorithm 2) with MaxHIG algorithm (Algorithm3). The following is implementation of algorithm MaxHIG on the example of Figure 2:

First we start with vertex 9 which has the maximum $U$, $U_9=4$, and add path $P(3\rightarrow 5\rightarrow 7\rightarrow 9)$ with cost 8 to the tree $Tree$. In the next step, vertex 8 with $U_8=3$ and cost 1 will be added to the tree. Then, the path $P(3\rightarrow 6)$ with cost 1 is added to the tree. The final result cost of the $Tree $ would be 10.

In fact, these two greedy algorithms are helpful for two different types of trees with different features. The first category that is solved with MinHIG are graphs, which in their optimal Steiner tree the basic vertices with high amount hops, $U$,  are connected to the basic vertices with lower $U$. The latter that can be solved with MaxHIG are graphs that in their optimal Steiner tree, some of the basic vertices are connected to the path of other basic vertices. To obtain Steiner tree in a given arbitrary graph, we run both algorithms and the best answer is considered as the final answer (we will see in last section that in most cases the result by combining of these two algorithms is better than the algorithm by Voss in \cite{14}). 
The complexity analysis of presented algorithms is as follows:\\
-$O(QEH)$ is needed to find the shortest path between basic vertices. \\
-$O(ELogE)$ is need for Prim algorithm implementation.\\
-$O(QV)$ is need for adding a basic vertex at each time and the comparison to all  vertices inside $Tree$.\\
\section{Non Root Based Insertion (NRBI) Algorithm}
\label{}
In this part, first we give an informal overview of  NRBI algorithm and then we present the analysis in details.
The algorithm uses generalized idea of both Prime and Kruskal algorithm. We use new variable $itr_v$ (Algorithm 4) to every basic vertex $v$ as time of entrance to the set $G$. Again here We assume that $itr_{root}$ is zero at first and then the first basic vertex $v$ entered to set $G$ after root has $itr_v=1$.
In phase 2 of NRBI algorithm, basic vertices are sorted in descending order due to their $iter$ amount. Then, they would be added to the $Tree$ similar to idea of Kruskal algorithm. In the Voss's algorithm, the final tree is a subset of set $G$ constructed in the first phase. In MinHIG and MaxHIG algorithms, set $G$ is completely neglected and the final tree is created only based on $U$ values. In this algorithm, as we will see further, set $G$ is used as an auxiliary set. When vertex $v$ is added to the tree, there will be two cases:

1- Connecting vertex $v$ to vertices with less $itr$.

2- Connecting vertex $v$ to vertices, which will be added to $G$ after vertex $v$.

In the former one, the result is clear and vertex $v$ should be connected to same vertices that it was connected to in $G$. If the latter, result will not be clear. First we have to find all vertices with more $itr$ than $v$ to find the needed Steiner vertices and add them to the $Tree$ and then we add vertex $v$.
Suppose all vertices except basic vertex $v$ have been added in the optimal tree and now we have to add the last remaining vertex $v$ to the Steiner tree. For similar conditions without loss of generality, we assume that the tree is two pieces. 
First one contains all vertices with $itr$ less than $k$ (this sub-tree may be a sub-forest and consists of many disconnected sub-trees) and second one is optimal sub-tree containing all vertices with $itr$ value more than $k$. Our goal is to connect $v$ to the first or second set in the best way. Each time, vertex $v$ is connected to one of the two pieces until we get the whole tree of all vertices especially vertex $v$. The optimal sub-tree of the first set of the vertices with $itr$ less than $k$ is the optimal sub-tree generated by the set $S$ containing all vertices added before $v$ to $S$. 

\bigskip
\begin{algorithm}[H]
\SetKwInOut{Input}{input}\SetKwInOut{Output}{output}
\Input{An undirected graph.}
\Output{U and itr values.}
{\bf Step 1}. {\bf Initialization} $G=(\{root\},\emptyset)$,

\While{$Q\not\subseteq S$}{

{\bf Step 2}. {\bf Find} vertices $u^*\in V_G$ and $v^*\in V_T$ and path P(u*,v*), where 

$U_v^* = U_u^* +|P(u^*,v^*)|\leq H$,

$d_{u^* v^*}=min⁡\{d_{uv} |u\leq V_G ,v^*\in V_T\}$;

{\bf Step 3}. {\bf Add} the vertices and edges of path $P(u^*,v^*)$ to G,

{\bf update} U values of the vertices of path $P(u^*,v^*)$,

{\bf Save} $itr[Basic Vertex\hspace{0.2cm} v \in P(u^*,v^* )]$.
}

\caption{NRBI (Phase 1)}
\end{algorithm}

\bigskip
\begin{algorithm}[H]
\SetKwInOut{Input}{input}\SetKwInOut{Output}{output}
\Input{$U_i$ and itr. }
\Output{An efficient feasible solution.}
{\bf Initialization} $Tree=\{\emptyset\}$

\For{Maximum $Itr_v$ to 1}{

{\bf Find} $P(v,u^* )$ where 

$u^*\in Tree$,

$U_u^*+|P(u^*, v )|\leq U_v$;

\If {$Cost(P(v,u^* ))<Cost(P(v,u^* )\in G)$}

{{\bf Add} all vertices and edges of $P(v,u^* )$ to Tree.}

\Else{ {\bf Add} all vertices and edges of $P(v,u^* )\in G$ to Tree.}
}
\caption{NRBI (Phase 2)}
\end{algorithm}
\bigskip
Now we want to find the most optimal connection of the vertex $v$ to the tree. So, in the second step, we try to make the best connections for optimal subtree of $v$. Since having $U$ values generated from the first phase of the generalized of prim algorithm, we can guarantee that the hop constraint is not violated. In the second part of the algorithm, by inspiring the idea of Kruskal algorithm, we allow to connect each vertex to other vertices without violating $U$ (Algorithm 5).
\paragraph{The analysis of NRBI algorithm}

Here we analyze our algorithm in details. We show that the output is tree and the constructed set doesn't include any cycle or cross. All basic vertices in $Q$ are connected together with $Q-1$ paths by idea of Kruskal algorithm. Because all $U$ values generated in the phase 1 are connected by this precondition that each vertex can only be connected to the vertex with less $U$, so no cycle would be created. 

Since, in this problem we consider adding path instead of edge due to of the idea of Kruskal algorithm, we should show that no cross is gonna happened either.
In NRBI algorithm, we connect vertices based on their $itr$ in descending order. The cross happens when some vertices of some paths are connected to each other. Since we know that paths that are constructed by vertices with more $itr$. These vertices are the only ones capable of changing $U$ values of vertices that are produced before in path by lower $itr$ and reverse of it is not possible. Thus cross wont happen.
In the second phase of the algorithm, when we add vertices with more $itr$ to the forest respectively, we assure that during choosing a shortest path for each basic vertex, no cross will be created (No changes can happen in previous paths of the forest).
To find the closest vertex to the current $itr$, distances from all basic and Steiner vertices that have been added to the $Tree$ before are calculated. Thus, there wont be any possibility of creating the cross. If we assume that cross is created, so the vertex at the intersection has been added to the $Tree$ before, then the vertex with current $itr$ was closer to the vertex of the cross and this is in contrast to our assumption that we have connected the current vertex to the closest vertex with respect to hop constraint. Therefore, there is no cross in graph and the result is tree.

Time complexity of this algorithm in the first part is similar to the previous presented algorithms, equal to $O(ELogE)$. In the second part in tree construction, all vertices are connected to each other in the forest. Since for every basic vertex when we add it, it will be compared to all the Steiner and basic vertices in the tree, thus the running time is $O(QV)$. Before, in the generalized Bellman-Ford, all paths from vertices to each other were calculated in time $O(QEH)$ and stored in the table. The only cost that we nee to calculate to find the path is the time of adding selected path from the table for every vertex in $Q$ which is at most $H$. Therefor, the time complexity would be $O(QH)$ and the total time of last part of the algorithm is $O(QV$).
%

%
\begin{figure}[h!]
\centering
\includegraphics[scale=0.25]{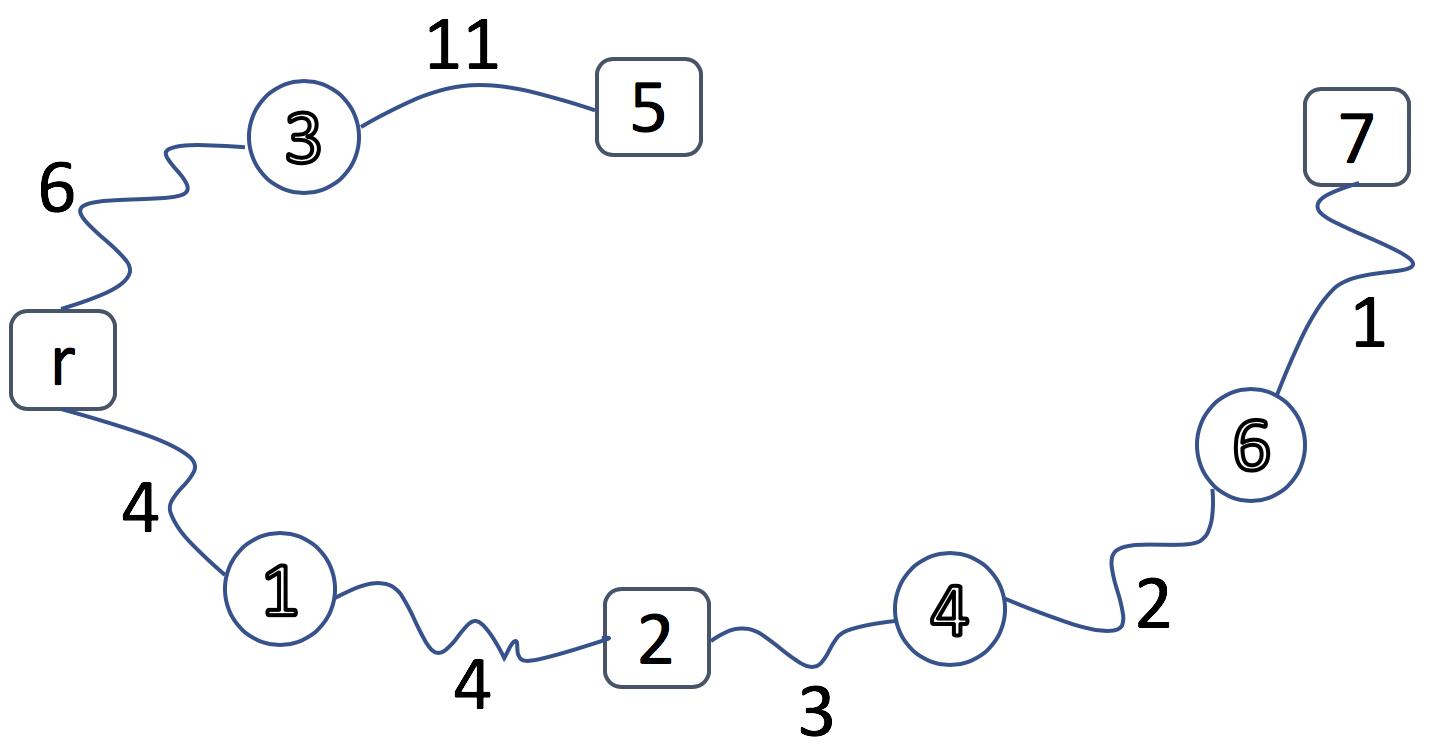}
\caption{An illustrative example to build HCST with 4 mendatory vertices and 4 Steiner vertices}
\label{A}
\end{figure}
Figure 3 presents an example that if we apply all previous algorithms on that we obtain 31 as the optimal Steiner tree. Though here is the implementation of the NRBI algorithm on the same graph:

{\bf First Step:}

First, paths $P(r\rightarrow 1\rightarrow 2)$ with cost 8 and basic vertex 2, $P(2\rightarrow 4\rightarrow 6\rightarrow 7)$ with cost of 6 and basic vertex 7, and then path $P(r\rightarrow 3\rightarrow 5)$ with cost 17 are added to set $G$ respectively. Now we see that set $G$ contains all basic vertices. 

The updated variables $itr$ and $U$ are as follows,

$itr_7=2, itr_5=3, itr_2=1,$ and $itr_0= 0$,

$U_5=2, U_2=2, U_7=5,$ and $U_0 = 0$ 

{\bf Second Step:}

So, we added basic vertices 0,2,5,7 to the tree.
Starting with the basic vertex with maximum $itr$, we select vertex 5. The shortest path between vertex 5 and existed vertices $u^*$ in set $Tree$ with $U_u^*+|P(u^*,5)|\leq U_5$ is $P(r\rightarrow 3\rightarrow 5)$ . All edges and vertices of this path are added to the tree (This path is the same path of vertex 5 in set $G$). 

The next basic vertex is the one with itr = 2, which is vertex 7.
The best way to connect vertex 7 to the one of the vertices in the set $Tree$ is path $P(2\rightarrow 4\rightarrow 6\rightarrow 7)$ with cost 6. For next vertex with $itr=1$, vertex 2, the best way to connect it to one of the vertices of the tree is $P(2\rightarrow 3)$ with cost 7. Although the best path for 2 in $G$ is $P(r\rightarrow 1\rightarrow 2)$ with cost 8, we add vertex 2 with cost 7. The optimal Steiner tree with cost 30 is obtained. 

Note, the tree at the beginning of the second phase is not connected same as idea of Kruskal algorithm when we add edges to find the shortest path, but with adding $Q-1$ paths, at the end the Steiner tree would be a connected tree.
\section{Computation}
\label{}
In order to assess the performance of all proposed algorithms, we used instances $\{c, d\}$n, $n\in \{5,10,15,20\}$ randomly chosen from the OR-Library\footnote{
http://people.brunel.ac.uk/mastjjb/jeb/orlib/steininfo.html}[6].

The size of these instances for ST problem have been defined between 500 and 1000 nodes from 625 up to 25000 edges (see \cite{105} for more details about these instances). 

In general, ST instances generate HCST input graphs in this way that ST files provide edge-costs and Steiner vertices so that we select 200 and 300 basic vertices due to random vectors. All algorithms have been implemented in C++ \footnote{https://github.com/Farzaneh9696/HC-Steiner} and run on all instances, sparse graphs $\{c5,c10,d5,d10\}$ and dense graphs $\{c15,c20,d15,d20\}$.

For the sake of quality of all algorithms for every sample library, we generate number of vectors equal to 100 times of the hop limitation $H$ to specify different group of basic vertices. For example, to implement the algorithms on instances for $H=3$, we generate 300 vectors to create different groups of basic vertices. 

All algorithms have been implemented on every vector on every instances. Therefore, the number of implementation for every algorithm on specified hop is equal to $100 H \times 16$.
We compare all of our algorithms with algorithm in \cite{14} (Named in the table Voss). In the tables, MinHIG algorithm, MaxHIG algorithm, and NRBI algorithm are represented as MinH, MaxH, and NRBI respectively. We also combine first two algorithms, MinHIG and MaxHIG, and show the results of this combination as MM in our tables.

Tables 1 to 8 compare results of algorithms as follows:

First algoritm vs second algorithm: {\bf FOS,\hspace{0.2cm} SFOS,\hspace{0.2cm} SOF, and\hspace{0.2cm} SSOF}

{ \bf FOS} = represents number of times that results of first algorithm is better than the second one.

{ \bf SFOS} = sum of the cost amount that the first algorithm is less than the second one. 

{\bf SOF}= represents the number of times that results of the second algorithm is better than the first one.

{\bf SSOF}= sum of the cost amount that the second algorithm is less than the first one.

In the following, we show that how we compare results of every two algorithms in every row of the table: 

For example:  Voss vs MinH: 6\hspace{0.2cm} 20 \hspace{0.2cm} 224\hspace{0.2cm} 2739 means that in this comparison the Voss's algorithm is 6 times better than MinH and the sum of the difference of their cost is 20. MinH is 224 times better than Voss and the sum of the difference of their cost is 2739. 

Tests are based on the benchmarks c and d with 200 and 300 basic vertices for 3, 5, 7, and 10 hop. Tables 1 to 4 present our computational results on sparse sets $\{c5,d5\}$ and $\{c10,d10\}$ with 200 and 300 basic vertices, respectively. We focus on NRBI and MM to show that they are the best ones among all five algorithms, but all algorithms are compared two each other one by one and results are provided in tables 1 to 8.

For $H=3$, in 8.54$\%$ of 300 vectors, NRBI is better than Voss algorithm. The cost difference is 69.12 in average. In the rest percent of the vectors, the cost calculation for both Voss and NRBI are the same, and in no cases Voss could get better than NRBI. For this hop limitation, We also saw MM is better than Voss with cost difference 49.37 in 5.79$\%$ of 300 vectors and in 2.5$\%$ vectors, Voss is better with cost difference 21.62 . In the remaining ones, two algorithms found the same cost. 

For $H=5$, NRBI in 67.37$\%$ of 500 vectors is better compared to the Voss's algorithm, with cost difference 12940.11 in average. In the remaining percent of the vectors the cost calculations for both Voss and NRBI are equal, and in no cases Voss was seen better than NRBI. In this hop, we also found that in 52.27$\%$ of 500 vectors, MM is better than Voss with cost difference 5879. In 14.2$\%$ vectors Voss is better with cost difference 638 and in the remaining ones two algorithms found the result with similar cost. 

For $H=7$, NRBI in 99.71$\%$ of 700 vectors is better than Voss with cost difference 34029.62 in average and for the rest of the vectors, the minimum costs  found by Voss and NRBI are equal.  In no cases Voss is better than NRBI. In this hop, in 63.14$\%$ of 700 vectors MM is better than Voss with cost difference 9040.12 and in 34.30$\%$ with cost 3604.37 vector Voss is better. The remaining vectors, two algorithms calculated same cost.
\begin{table}[H]
\begin{center}
\begin{adjustbox}{width=\textwidth,totalheight=\textheight,keepaspectratio}
\begin{tabular}{cc c c c|c c c c c}
\caption{Steiner results on c5 and d5 instances with 200 basic vertices}
\\ 
Methods Comparison&\multicolumn{4}{c}{c5} & \multicolumn{4}{|c}{d5}&Number of Hop \\ 
\hline 
&FOS &SFOS &SOF &SSOF &FOS &SFOS &SOF &SSOF& \\ 
\cline{2-9}
Voss vs MinH: &0 &0 &0 &0 &0 &0 &0 &0 & \\ 
Voss vs MaxH: &0 &0 &0 &0 &0 &0 &0 &0 & \\ 
Voss vs NRBI: &0 &0 &0 &0 &0 &0 &0 &0 &\\
Voss vs MM: &0 &0 &0 &0 &0 &0 &0 &0 &\\
MinH vs MaxH: &0 &0 &0 &0 &0 &0 &0 &0 &H=3\\
MinH vs NRBI: &0 &0 &0 &0 &0 &0 &0 &0 &\\
MinH vs MM: &0 &0 &0 &0 &0 &0 &0 &0 &\\
MaxH vs NRBI: &0 &0 &0 &0 &0 &0 &0 &0 &\\
MaxH vs MM: &0 &0 &0 &0 &0 &0 &0 &0 &\\
NRBI vs MM: &0 &0 &0 &0 &0 &0 &0 &0 &\\
\hline
Voss vs MinH: &0 &0 &96 &638 &6 &20 &224 &2739 &\\
Voss vs MaxH: &0 &0 &96 &638 &22 &118 &219 &2655 &\\
Voss vs NRBI: &0 &0 &96 &638 &0 &0 &231 &2805 &\\
Voss vs MM: &0 &0 &96 &638 &4 &16 &224 &2741 &\\
MinH vs MaxH: &0 &0 &0 &0 &33 &188 &3 &6 &H=5\\
MinH vs NRBI: &0 &0 &0 &0 &0 &0 &21 &86 &\\
MinH vs MM: &0 &0 &0 &0 &0 &0 &3 &6 &\\
MaxH vs NRBI: &0 &0 &0 &0 &0 &0 &49 &268 &\\
MaxH vs MM: &0 &0 &0 &0 &0 &0 &33 &188 &\\
NRBI vs MM: &0 &0 &0 &0 &18 &80 &0 &0 &\\
\hline
Voss vs MinH: &222 &1689 &448 &5801 &281 &3297 &399 &5753 &\\
Voss vs MaxH: &382 &4682 &302 &3422 &441 &7373 &240 &2814 &\\
Voss vs NRBI: &0 &0 &691 &14835 &0 &0 &697 &19983 &\\
Voss vs MM: &201 &1379 &470 &6515 &248 &2714 &428 &6258 &\\
MinH vs MaxH: &524 &6396 &155 &1024 &512 &8103 &115 &1088 &H=7\\
MinH vs NRBI: &24 &68 &662 &10791 &11 &26 &683 &17553 &\\
MinH vs MM: &0 &0 &155 &1024 &0 &0 &115 &1088 &\\
MaxH vs NRBI: &9 &48 &684 &16143 &0 &0 &698 &24542 &\\
MaxH vs MM: &0 &0 &524 &6396 &0 &0 &512 &8103 &\\
NRBI vs MM: &651 &9811 &31 &112 &682 &16465 &11 &26 &\\
\hline
Voss vs MinH: &296 &5671 &697 &18648 &563 &13046 &422 &8136 &\\
Voss vs MaxH: &762 &27232 &231 &4129 &921 &42692 &73 &1065 &\\
Voss vs NRBI: &0 &0 &1000 &49505 &0 &0 &1000 &46771 &\\
Voss vs MM: &279 &4712 &715 &19223 &553 &12259 &432 &8400 &\\
MinH vs MaxH: &894 &37614 &101 &1534 &902 &37768 &92 &1051 &H=10\\
MinH vs NRBI: &13 &75 &986 &36603 &1 &6 &999 &51687 &\\
MinH vs MM: &0 &0 &101 &1534 &0 &0 &92 &1051 &\\
MaxH vs NRBI: &1 &7 &999 &72615 &0 &0 &1000 &88398 &\\
MaxH vs MM: &0 &0 &894 &37614 &0 &0 &902 &37768 &\\
NRBI vs MM: &985 &35076 &14 &82 &999 &50636 &1 &6 &\\
\hline
\end{tabular}
\end{adjustbox}
\end{center}
\end{table}
For $H=10$, in 99.75$\%$ of 1000 vectors, NRBI could find a better solution than Voss  with cost difference 9683 in average. In the remaining percent of the vectors Voss and NRBI are equal. In 39.95$\%$ of 1000 vectors, MM is better than Voss with cost difference 38545.75. Also, Voss is better in 58.67$\%$ vectors with cost difference 1789.75.Remaining had same results.
\begin{table}[H]
\begin{center}
\begin{adjustbox}{width=\textwidth,totalheight=\textheight,keepaspectratio}
\begin{tabular}{cc c c c|c c c c c}
\\ 
Methods Comparison&\multicolumn{4}{c}{c5} & \multicolumn{4}{|c}{d5}&Number of Hop \\ 
\hline 
&FOS &SFOS &SOF &SSOF &FOS &SFOS &SOF &SSOF& \\ 
\cline{2-9}
Voss vs MinH: &0 &0 &0 &0 &0 &0 &0 &0 &\\
Voss vs MaxH: &0 &0 &0 &0 &0 &0 &0 &0 &\\
Voss vs NRBI: &0 &0 &0 &0 &0 &0 &0 &0 &\\
Voss vs MM: &0 &0 &0 &0 &0 &0 &0 &0 &\\
MinH vs MaxH: &0 &0 &0 &0 &0 &0 &0 &0 &H=3\\
MinH vs NRBI: &0 &0 &0 &0 &0 &0 &0 &0 &\\
MinH vs MM: &0 &0 &0 &0 &0 &0 &0 &0 &\\
MaxH vs NRBI: &0 &0 &0 &0 &0 &0 &0 &0 &\\
MaxH vs MM: &0 &0 &0 &0 &0 &0 &0 &0 &\\
NRBI vs MM: &0 &0 &0 &0 &0 &0 &0 &0 &\\
\hline
Voss vs MinH: &94 &518 &46 &264 &7 &22 &52 &459 &\\
Voss vs MaxH: &160 &670 &48 &269 &64 &230 &250 &2929 &\\
Voss vs NRBI: &0 &0 &116 &540 &0 &0 &253 &3107 &\\
Voss vs MM: &74 &365 &54 &304 &1 &3 &252 &2950 &\\
MinH vs MaxH: &109 &340 &39 &193 &67 &248 &220 &2510 &H=5\\
MinH vs NRBI: &6 &36 &152 &830 &0 &0 &222 &2670 &\\
MinH vs MM: &0 &0 &39 &193 &0 &0 &220 &2510 &\\
MaxH vs NRBI: &7 &34 &217 &975 &0 &0 &112 &408 &\\
MaxH vs MM: &0 &0 &109 &340 &0 &0 &67 &248 &\\
NRBI vs MM: &130 &642 &7 &41 &46 &160 &0 &0 &\\ 
\hline
Voss vs MinH: &291 &2909 &387 &5108 &171 &1623 &513 &8615 &\\
Voss vs MaxH: &358 &5280 &317 &4029 &350 &5619 &340 &5118 &\\
Voss vs NRBI: &0 &0 &698 &17979 &0 &0 &698 &19602 &\\
Voss vs MM: &221 &1922 &453 &6487 &134 &1080 &548 &9751 &\\
MinH vs MaxH: &424 &5816 &247 &2366 &515 &9172 &170 &1679 &\\
MinH vs NRBI: &5 &13 &693 &15793 &19 &80 &673 &12690 &H=7\\
MinH vs MM: &0 &0 &247 &2366 &0 &0 &170 &1679 &\\
MaxH vs NRBRI: &5 &13 &693 &19243 &5 &26 &693 &20129 &\\
MaxH vs MM: &0 &0 &424 &5816 &0 &0 &515 &9172 &\\
NRBI vs MM: &687 &13436 &9 &22 &666 &11037 &24 &106 &\\ 
\hline
Voss vs MinH: &277 &5560 &718 &25206 &380 &7035 &598 &14452 &\\
Voss vs MaxH: &711 &27168 &279 &7472 &784 &31241 &207 &3655 &\\
Voss vs NRBI: &0 &0 &1000 &64750 &0 &0 &1000 &62950 &\\
Voss vs MM: &270 &5223 &725 &25720 &352 &6131 &628 &15403 &\\
MinH vs MaxH: &925 &40193 &71 &851 &866 &36858 &126 &1855 &H=10\\
MinH vs NRBI: &14 &92 &984 &45196 &0 &0 &999 &55533 &\\
MinH vs MM: &0 &0 &71 &851 &0 &0 &126 &1855 &\\
MaxH vs NRBI: &1 &1 &999 &84447 &0 &0 &1000 &90536 &\\
MaxH vs MM: &0 &0 &925 &40193 &0 &0 &866 &36858 &\\
NRBI vs MM: &983 &44346 &15 &93 &999 &53678 &0 &0 &\\
\hline
\end{tabular}
\end{adjustbox}
\end{center}
\caption{Steiner results on c5 and d5 instances with 300 basic vertices}
\end{table}
\begin{table}[H]
\begin{center}
\begin{adjustbox}{width=\textwidth,totalheight=\textheight,keepaspectratio}
\begin{tabular}{cc c c c|c c c c c}
\\ 
Methods Comparison&\multicolumn{4}{c}{c10} & \multicolumn{4}{|c}{d10}&Number of Hop \\ 
\hline 
&FOS &SFOS &SOF &SSOF &FOS &SFOS &SOF &SSOF& \\ 
\cline{2-9}
Voss vs MinH: &21 &70 &18 &64 &0 &0 &0 &0 &\\ 
Voss vs MaxH: &29 &109 &32 &105 &0 &0 &0 &0 &\\
Voss vs NRBI: &0 &0 &60 &166 &0 &0 &0 &0 &\\
Voss vs MM: &16 &46 &37 &120 &0 &0 &0 &0 &\\
MinH vs MaxH: &18 &78 &24 &80 &0 &0 &0 &0 &H=3\\
MinH vs NRBI: &0 &0 &63 &172 &0 &0 &0 &0 &\\
MinH vs MM: &0 &0 &24 &80 &0 &0 &0 &0 &\\
MaxH vs NRBI: &0 &0 &57 &170 &0 &0 &0 &0 &\\
MaxH vs MM: &0 &0 &18 &78 &0 &0 &0 &0 &\\
NRBI vs MM: &39 &92 &0 &0 &0 &0 &0 &0 &\\
\hline
Voss vs MinH: &39 &327 &456 &14096 &264 &3296 &222 &2386 &\\
Voss vs MaxH: &175 &2562 &317 &6787 &233 &2722 &257 &2533 &\\
Voss vs NRBI: &0 &0 &500 &28924 &0 &0 &500 &12442 &\\
Voss vs MM: &28 &200 &468 &14740 &196 &2019 &293 &3284 &\\
MinH vs MaxH: &416 &10315 &76 &771 &215 &1454 &247 &2175 &H=5\\
MinH vs NRBI: &1 &3 &498 &15158 &0 &0 &498 &13352 &\\
MinH vs MM: &0 &0 &76 &771 &0 &0 &247 &2175 &\\
MaxH vs NRBI: &0 &0 &499 &24699 &1 &1 &498 &12632 &\\
MaxH vs MM: &0 &0 &416 &10315 &0 &0 &215 &1454 &\\
NRBI vs MM: &497 &14387 &1 &3 &496 &11178 &1 &1 &\\
\hline
Voss vs MinH: &375 &6439 &309 &4436 &497 &17432 &195 &5241 &\\
Voss vs MaxH: &680 &34162 &19 &150 &347 &9000 &344 &9944 &\\
Voss vs NRBI: &0 &0 &700 &23774 &0 &0 &700 &52782 &\\
Voss vs MM: &374 &6287 &310 &4450 &311 &7199 &378 &11278 &\\
MinH vs MaxH: &682 &32175 &17 &166 &178 &3135 &517 &16270 &H=7\\
MinH vs NRBI: &1 &3 &699 &25780 &0 &0 &700 &64973 &\\
MinH vs MM: &0 &0 &17 &166 &0 &0 &517 &16270 &\\
MaxH vs NRBI: &0 &0 &700 &57786 &0 &0 &700 &51838 &\\
MaxH vs MM: &0 &0 &682 &32175 &0 &0 &178 &3135 &\\
NRBI vs MM: &699 &25614 &1 &3 &700 &48703 &0 &0 &\\ 
\hline
Voss vs MinH: &834 &14577 &149 &1182 &986 &56132 &12 &150 &\\
Voss vs MaxH: &1000 &97700 &0 &0 &999 &101492 &1 &23 &\\
Voss vs NRBI: &0 &0 &982 &13866 &0 &0 &1000 &39112 &\\
Voss vs MM: &834 &14577 &149 &1182 &985 &54496 &13 &173 &\\
MinH vs MaxH: &999 &84305 &0 &0 &903 &47146 &93 &1659 &H=10\\
MinH vs NRBI: &0 &0 &997 &27261 &0 &0 &1000 &95094 &\\
MinH vs MM: &0 &0 &0 &0 &0 &0 &93 &1659 &\\
MaxH vs NRBI: &0 &0 &1000 &111566 &0 &0 &1000 &140581 &\\
MaxH vs MM: &0 &0 &999 &84305 &0 &0 &903 &47146 &\\
NRBI vs MM: &997 &27261 &0 &0 &1000 &93435 &0 &0 &\\
\hline
\end{tabular}
\end{adjustbox}
\end{center}
\caption{Steiner results on c10 and d10 instances with 200 basic vertices}
\end{table}
Tables 5 to 8 present our computational results on dense sets $\{c15,d15\}$ and $\{c20,d20\}$ with 200 and 300 basic vertices, respectively. When $H=3$, NRBI is better than Voss in 100$\%$ of 300 vectors. The cost difference is 9683 in average. In the remaining part of the vectors Voss and NRBI both have equal result, and in no cases Voss is better than NRBI. In this hop, in 58.83$\%$ of 300 vectors,  MM is better than Voss with cost difference 2407.12 and in 37.16$\%$ of them Voss gets better with cost 797. For rest of the vectors they calculate same cost.
\begin{table}[H]
\begin{center}
\begin{adjustbox}{width=\textwidth,totalheight=\textheight,keepaspectratio}
\begin{tabular}{cc c c c|c c c c c}
\\ 
Method Comparison&\multicolumn{4}{c}{c10} & \multicolumn{4}{|c}{d10}&Number of Hop \\ 
\hline 
&FOS &SFOS &SOF &SSOF &FOS &SFOS &SOF &SSOF& \\ 
\cline{2-9}
Voss vs MinH: &68 &266 &51 &79 &0 &0 &0 &0 &\\ 
Voss vs MaxH: &87 &311 &91 &247 &0 &0 &0 &0 &\\
Voss vs NRBI: &0 &0 &145 &387 &0 &0 &0 &0 &\\
Voss vs MM: &44 &127 &102 &275 &0 &0 &0 &0 &\\
MinH vs MaxH: &59 &212 &93 &335 &0 &0 &0 &0 &H=3\\
MinH vs NRBI: &0 &0 &154 &574 &0 &0 &0 &0 &\\
MinH vs MM: &0 &0 &93 &335 &0 &0 &0 &0 &\\
MaxH vs NRBI: &0 &0 &116 &451 &0 &0 &0 &0 &\\
MaxH vs MM: &0 &0 &59 &212 &0 &0 &0 &0 &\\
NRBI vs MM: &81 &239 &0 &0 &0 &0 &0 &0 &\\
\hline
Voss vs MinH: &20 &125 &473 &20236 &278 &3889 &210 &2631 &\\
Voss vs MaxH: &181 &3499 &314 &7960 &281 &3901 &207 &2101 &\\
Voss vs NRBI: &0 &0 &500 &36101 &0 &0 &499 &18964 &\\
Voss vs MM: &16 &101 &479 &20437 &218 &2686 &272 &3302 &\\
MinH vs MaxH: &472 &15875 &24 &225 &253 &2416 &218 &1874 &H=5\\
MinH vs NRBI: &1 &1 &497 &15991 &0 &0 &500 &20222 &\\
MinH vs MM: &0 &0 &24 &225 &0 &0 &218 &1874 &\\
MaxH vs NRBI: &0 &0 &500 &31640 &0 &0 &500 &20764 &\\
MaxH vs MM: &0 &0 &472 &15875 &0 &0 &253 &2416 &\\
NRBI vs MM: &497 &15766 &1 &1 &500 &18348 &0 &0 &\\
\hline
Voss vs MinH: &194 &2423 &491 &10725 &260 &6758 &435 &15972 &\\
Voss vs MaxH: &694 &48742 &3 &69 &520 &22808 &176 &4618 &\\
Voss vs NRBI: &0 &0 &700 &31022 &0 &0 &700 &92260 &\\
Voss vs MM: &194 &2421 &491 &10730 &238 &5833 &458 &16852 &\\
MinH vs MaxH: &698 &56982 &2 &7 &591 &29209 &105 &1805 &H=7\\
MinH vs NRBI: &1 &5 &699 &22725 &0 &0 &700 &83046 &\\
MinH vs MM: &0 &0 &2 &7 &0 &0 &105 &1805 &\\
MaxH vs NRBI: &0 &0 &700 &79695 &0 &0 &700 &110450 &\\
MaxH vs MM: &0 &0 &698 &56982 &0 &0 &591 &29209 &\\
NRBI vs MM: &699 &22718 &1 &5 &700 &81241 &0 &0 &\\
\hline
Voss vs MinH : &560 &6496 &397 &3949 &861 &40240 &137 &2817 &\\
Voss vs MaxH : &1000 &127553 &0 &0 &996 &138991 &4 &119 &\\
Voss vs NRBI : &0 &0 &998 &20109 &0 &0 &1000 &68787 &\\
Voss vs MM : &560 &6496 &397 &3949 &861 &40024 &137 &2832 &\\
MinH vs MaxH : &1000 &125006 &0 &0 &989 &101680 &10 &231 &H=10\\
MinH vs NRBI: &1 &3 &998 &22659 &0 &0 &1000 &106210 &\\
MinH vs MM : &0 &0 &0 &0 &0 &0 &10 &231 &\\
MaxH vs NRBI: &0 &0 &1000 &147662 &0 &0 &1000 &207659 &\\
MaxH vs MM : &0 &0 &1000 &125006 &0 &0 &989 &101680 &\\
NRBI vs MM : &998 &22659 &1 &3 &1000 &105979 &0 &0 &\\
\hline
\end{tabular}
\end{adjustbox}
\end{center}
\caption{Steiner results on c10 and d10 instances with 300 basic vertices}
\end{table}
\begin{table}[H]
\begin{center}
\begin{adjustbox}{width=\textwidth,totalheight=\textheight,keepaspectratio}
\begin{tabular}{cc c c c|c c c c c}
\\ 
Methods Comparison&\multicolumn{4}{c}{c15} & \multicolumn{4}{|c}{d15}&Number of Hop \\ 
\hline 
&FOS &SFOS &SOF &SSOF &FOS &SFOS &SOF &SSOF& \\ 
\cline{2-9}
Voss vs MinH: &67 &677 &227 &4224 &39 &182 &250 &2380 &\\ 
Voss vs MaxH: &151 &2010 &137 &1738 &93 &578 &190 &1439 &\\
Voss vs NRBI: &0 &0 &300 &15814 &0 &0 &299 &5292 &\\
Voss vs MM: &61 &583 &231 &4389 &35 &169 &254 &2451 &\\
MinH vs MaxH: &255 &4078 &42 &259 &238 &1421 &20 &84 &H=3\\
MinH vs NRBI: &0 &0 &299 &12267 &1 &1 &293 &3095 &\\
MinH vs MM: &0 &0 &42 &259 &0 &0 &20 &84 &\\
MaxH vs NRBI: &0 &0 &300 &16086 &1 &3 &298 &4434 &\\
MaxH vs MM: &0 &0 &255 &4078 &0 &0 &238 &1421 &\\
NRBI vs MM: &299 &12008 &0 &0 &292 &3014 &2 &4 &\\
\hline
Voss vs MinH: &13 &141 &486 &20420 &5 &22 &495 &22888 &\\
Voss vs MaxH: &304 &5007 &187 &2488 &37 &221 &459 &11779 &\\
Voss vs NRBI: &0 &0 &500 &21862 &0 &0 &500 &21389 &\\
Voss vs MM: &306 &5019 &185 &2434 &41 &241 &455 &11297 &\\
MinH vs MaxH: &488 &22864 &8 &66 &440 &11810 &56 &502 &H=5\\
MinH vs NRBI: &0 &0 &500 &19343 &0 &0 &500 &32947 &\\
MinH vs MM: &0 &0 &8 &66 &0 &0 &56 &502 &\\
MaxH vs NRBI: &0 &0 &500 &42141 &0 &0 &500 &44255 &\\
MaxH vs MM: &0 &0 &488 &22864 &0 &0 &440 &11810 &\\
NRBI vs MM: &500 &19277 &0 &0 &500 &32445 &0 &0 &\\ 
\hline
Voss vs MinH: &0 &0 &700 &64212 &1 &3 &699 &44764 &\\
Voss vs MaxH: &97 &649 &593 &10549 &6 &20 &694 &21112 &\\
Voss vs NRBI: &0 &0 &699 &12872 &0 &0 &700 &11019 &\\
Voss vs MM: &97 &649 &593 &10549 &6 &20 &694 &21016 &\\
MinH vs MaxH: &700 &54312 &0 &0 &680 &23765 &18 &96 &H=7\\
MinH vs NRBI: &0 &0 &700 &22772 &0 &0 &700 &32111 &\\
MinH vs MM: &0 &0 &0 &0 &0 &0 &18 &96 &\\
MaxH vs NRBI: &0 &0 &700 &77084 &0 &0 &700 &55780 &\\
MaxH vs MM: &0 &0 &700 &54312 &0 &0 &680 &23765 &\\
NRBI vs MM: &700 &22772 &0 &0 &700 &32015 &0 &0 &\\ 
\hline
Voss vs MinH: &0 &0 &1000 &123718 &0 &0 &1000 &83140 &\\
Voss vs MaxH: &39 &171 &947 &18998 &0 &0 &1000 &34330 &\\
Voss vs NRBI: &0 &0 &980 &7150 &0 &0 &956 &5709 &\\
Voss vs MM: &39 &171 &947 &18998 &0 &0 &1000 &34322 &\\
MinH vs MaxH: &1000 &104891 &0 &0 &997 &48818 &3 &8 &H=10\\
MinH vs NRBI: &0 &0 &1000 &25977 &0 &0 &1000 &40039 &\\
MinH vs MM: &0 &0 &0 &0 &0 &0 &3 &8 &\\
MaxH vs NRBI: &0 &0 &1000 &130868 &0 &0 &1000 &88849 &\\
MaxH vs MM: &0 &0 &1000 &104891 &0 &0 &997 &48818 &\\
NRBI vs MM: &1000 &25977 &0 &0 &1000 &40031 &0 &0 &\\
\hline
\end{tabular}
\end{adjustbox}
\end{center}
\caption{Steiner results on c15 and d15 instances with 200 basic vertices}
\end{table}
For $H=5$, in 92.70$\%$ of 500 vectors NRBI result is better than what Voss got with cost difference 13012.25 in average. In the rest percent of the vectors, the min cost calculated by Voss and 
NRBI are same.  In no cases Voss is better than NRBI. For this hop, also in 77.12$\%$ of 500 vectors MM is better than Voss. The cost difference of them is 4101. In 20.8$\%$ of vectors Voss is better with cost difference 1657.62. For rest of the vectors two algorithms calculate same cost.
\begin{table}[H]
\begin{center}
\begin{adjustbox}{width=\textwidth,totalheight=\textheight,keepaspectratio}
\begin{tabular}{cc c c c|c c c c c}
\\ 
Methods Comparison&\multicolumn{4}{c}{c15} & \multicolumn{4}{|c}{d15}&Number of Hop \\ 
\hline 
&FOS &SFOS &SOF &SSOF &FOS &SFOS &SOF &SSOF& \\ 
\cline{2-9}
Voss vs MinH: &29 &233 &268 &7410 &116 &690 &179 &1768 &\\ 
Voss vs MaxH: &131 &2032 &162 &3205 &168 &1560 &119 &1071 &\\
Voss vs NRBI: &0 &0 &300 &22042 &0 &0 &300 &7822 &\\
Voss vs MM: &27 &219 &271 &7594 &112 &648 &181 &1856 &\\
MinH vs MaxH: &270 &6202 &25 &198 &238 &1697 &45 &130 &H=3\\
MinH vs NRBI: &0 &0 &300 &14865 &0 &0 &300 &6744 &\\
MinH vs MM: &0 &0 &25 &198 &0 &0 &45 &130 &\\
MaxH vs NRBI: &0 &0 &300 &20869 &0 &0 &300 &8311 &\\
MaxH vs MM: &0 &0 &270 &6202 &0 &0 &238 &1697 &\\
NRBI vs MM: &300 &14667 &0 &0 &300 &6614 &0 &0 &\\
\hline
Voss vs MinH: &9 &158 &490 &34526 &2 &24 &498 &34875 &\\
Voss vs MaxH: &367 &7289 &120 &1084 &63 &592 &434 &12094 &\\
Voss vs NRBI: &0 &0 &500 &25233 &0 &0 &500 &27998 &\\
Voss vs MM: &367 &7289 &120 &1067 &64 &600 &433 &11977 &\\
MinH vs MaxH: &499 &40590 &1 &17 &481 &23474 &17 &125 &H=5\\
MinH vs NRBI: &0 &0 &500 &19028 &0 &0 &500 &39500 &\\
MinH vs MM: &0 &0 &1 &17 &0 &0 &17 &125 &\\
MaxH vs NRBI: &0 &0 &500 &59601 &0 &0 &500 &62849 &\\
MaxH vs MM: &0 &0 &499 &40590 &0 &0 &481 &23474 &\\
NRBI vs MM: &500 &19011 &0 &0 &500 &39375 &0 &0 &\\
\hline
Voss vs MinH: &0 &0 &700 &78845 &0 &0 &700 &59576 &\\
Voss vs MaxH: &250 &2070 &421 &4763 &17 &124 &677 &19948 &\\
Voss vs NRBI: &0 &0 &700 &15325 &0 &0 &700 &18595 &\\
Voss vs MM: &250 &2070 &421 &4763 &17 &124 &677 &19940 &\\
MinH vs MaxH: &700 &76152 &0 &0 &697 &39760 &1 &8 &H=7\\
MinH vs NRBI: &0 &0 &700 &18018 &0 &0 &700 &38419 &\\
MinH vs MM: &0 &0 &0 &0 &0 &0 &1 &8 &\\
MaxH vs NRBI: &0 &0 &700 &94170 &0 &0 &700 &78171 &\\
MaxH vs MM: &0 &0 &700 &76152 &0 &0 &697 &39760 &\\
NRBI vs MM: &700 &18018 &0 &0 &700 &38411 &0 &0 &\\ 
\hline
Voss vs MinH: &0 &0 &1000 &145523 &0 &0 &1000 &114865 &\\
Voss vs MaxH: &90 &472 &892 &11882 &2 &4 &998 &36087 &\\
Voss vs NRBI: &0 &0 &992 &8653 &0 &0 &994 &9344 &\\
Voss vs MM: &90 &472 &892 &11882 &2 &4 &998 &36087 &\\
MinH vs MaxH: &1000 &134113 &0 &0 &1000 &78782 &0 &0 &H=10\\
MinH vs NRBI: &0 &0 &1000 &20063 &0 &0 &1000 &45427 &\\
MinH vs MM: &0 &0 &0 &0 &0 &0 &0 &0 &\\
MaxH vs NRBI: &0 &0 &1000 &154176 &0 &0 &1000 &124209 &\\
MaxH vs MM: &0 &0 &1000 &134113 &0 &0 &1000 &78782 &\\
NRBI vs MM: &1000 &20063 &0 &0 &1000 &45427 &0 &0 &\\
\hline
\end{tabular}
\end{adjustbox}
\end{center}
\caption{Steiner results on c15 and d15 instances with 300 basic vertices}
\end{table}
\begin{table}[H]
\begin{center}
\begin{adjustbox}{width=\textwidth,totalheight=\textheight,keepaspectratio}
\begin{tabular}{cc c c c|c c c c c}
\\ 
Methods Comparison&\multicolumn{4}{c}{c20} & \multicolumn{4}{|c}{d20}&Number of Hop \\ 
\hline 
&FOS &SFOS &SOF &SSOF &FOS &SFOS &SOF &SSOF& \\ 
\cline{2-9}
Voss vs MinH: &166 &706 &105 &412 &229 &1819 &57 &288 &\\ 
Voss vs MaxH: &286 &3298 &9 &30 &293 &4831 &6 &29 &\\
Voss vs NRBI: &0 &0 &300 &3484 &0 &0 &300 &6621 &\\
Voss vs MM: &164 &696 &107 &425 &229 &1808 &57 &293 &\\
MinH vs MaxH: &289 &2997 &9 &23 &289 &3287 &8 &16 &H=3\\
MinH vs NRBI: &0 &0 &299 &3778 &0 &0 &300 &8152 &\\
MinH vs MM: &0 &0 &9 &23 &0 &0 &8 &16 &\\
MaxH vs NRBI: &0 &0 &300 &6752 &0 &0 &300 &11423 &\\
MaxH vs MM: &0 &0 &289 &2997 &0 &0 &289 &3287 &\\
NRBI vs MM: &299 &3755 &0 &0 &300 &8136 &0 &0 &\\ 
\hline
Voss vs MinH: &0 &0 &500 &20868 &0 &0 &500 &21531 &\\
Voss vs MaxH: &14 &26 &475 &3163 &5 &17 &493 &5853 &\\
Voss vs NRBI: &0 &0 &380 &987 &0 &0 &493 &2560 &\\
Voss vs MM: &14 &26 &475 &3163 &5 &17 &493 &5852 &\\
MinH vs MaxH: &500 &17731 &0 &0 &499 &15696 &1 &1 &H=5\\
MinH vs NRBI: &0 &0 &499 &4124 &0 &0 &500 &8396 &\\
MinH vs MM: &0 &0 &0 &0 &0 &0 &1 &1 &\\
MaxH vs NRBI: &0 &0 &500 &21855 &0 &0 &500 &24091 &\\
MaxH vs MM: &0 &0 &500 &17731 &0 &0 &499 &15696 &\\
NRBI vs MM: &499 &4124 &0 &0 &500 &8395 &0 &0 &\\ 
\hline
Voss vs MinH: &0 &0 &700 &34709 &0 &0 &700 &39599 &\\
Voss vs MaxH: &0 &0 &694 &5152 &1 &2 &699 &8067 &\\
Voss vs NRBI: &0 &0 &231 &275 &0 &0 &596 &1834 &\\
Voss vs MM: &0 &0 &694 &5152 &1 &2 &699 &8067 &\\
MinH vs MaxH: &700 &29557 &0 &0 &700 &31534 &0 &0 &H=7\\
MinH vs NRBI: &0 &0 &696 &5427 &0 &0 &700 &9899 &\\
MinH vs MM: &0 &0 &0 &0 &0 &0 &0 &0 &\\
MaxH vs NRBI: &0 &0 &700 &34984 &0 &0 &700 &41433 &\\
MaxH vs MM: &0 &0 &700 &29557 &0 &0 &700 &31534 &\\
NRBI vs MM: &696 &5427 &0 &0 &700 &9899 &0 &0 &\\ 
\hline
Voss vs MinH: &0 &0 &1000 &49957 &0 &0 &1000 &64645 &\\
Voss vs MaxH: &1 &1 &990 &7125 &0 &0 &1000 &13254 &\\
Voss vs NRBI: &0 &0 &263 &307 &0 &0 &325 &465 &\\
Voss vs MM: &1 &1 &990 &7125 &0 &0 &1000 &13254 &\\
MinH vs MaxH: &1000 &42833 &0 &0 &1000 &51391 &0 &0 &H=10\\
MinH vs NRBI: &0 &0 &994 &7431 &0 &0 &1000 &13719 &\\
MinH vs MM: &0 &0 &0 &0 &0 &0 &0 &0 &\\
MaxH vs NRBI: &0 &0 &1000 &50264 &0 &0 &1000 &65110 &\\
MaxH vs MM: &0 &0 &1000 &42833 &0 &0 &1000 &51391 &\\
NRBI vs MM: &994 &7431 &0 &0 &1000 &13719 &0 &0 &\\
\hline
\end{tabular}
\end{adjustbox}
\end{center}
\caption{Steiner results on c20 and d20 instances with 200 basic vertices}
\end{table}
For $H=7$, NRBI in 75.50$\%$ of 700 vectors is better than Voss with cost difference 7608.3 in average. In the remained percent of the vectors the minimum costs calculated by both Voss and NRBI are equal. In no cases Voss is better than NRBI. For this hop, in 91.6$\%$ of 700 vectors MM is better than Voss with cost difference 10210.5 and in 5.32$\%$ vectors Voss is better with cost difference 1090.37.
\begin{table}[H]
\begin{center}
\begin{adjustbox}{width=\textwidth,totalheight=\textheight,keepaspectratio}
\begin{tabular}{cc c c c|c c c c c}
\\ 
Methods Comparison&\multicolumn{4}{c}{c20} & \multicolumn{4}{|c}{d20}&Number of Hop \\ 
\hline 
&FOS &SFOS &SOF &SSOF &FOS &SFOS &SOF &SSOF& \\ 
\cline{2-9}
Voss vs MinH: &59 &299 &228 &1763 &206 &1875 &83 &481 &\\ 
Voss vs MaxH: &288 &4347 &8 &27 &290 &5431 &8 &51 &\\
Voss vs NRBI: &0 &0 &300 &5951 &0 &0 &300 &10438 &\\
Voss vs MM: &59 &293 &228 &1766 &205 &1860 &83 &483 &\\
MinH vs MaxH: &296 &5793 &3 &9 &284 &4003 &6 &17 &H=3\\
MinH vs NRBI: &1 &1 &299 &4488 &0 &0 &300 &11832 &\\
MinH vs MM: &0 &0 &3 &9 &0 &0 &6 &17 &\\
MaxH vs NRBI: &0 &0 &300 &10271 &0 &0 &300 &15818 &\\
MaxH vs MM : &0 &0 &296 &5793 &0 &0 &284 &4003 &\\
NRBI vs MM: &299 &4479 &1 &1 &300 &11815 &0 &0 &\\ 
\hline
Voss vs MinH: &0 &0 &500 &29162 &0 &0 &500 &27343 &\\
Voss vs MaxH: &23 &31 &440 &1950 &12 &38 &484 &5848 &\\
Voss vs NRBI: &0 &0 &339 &684 &0 &0 &496 &3386 &\\
Voss vs MM: &23 &31 &440 &1950 &12 &38 &484 &5848 &\\
MinH vs MaxH: &500 &27243 &0 &0 &500 &21533 &0 &0 &H=5\\
MinH vs NRBI: &0 &0 &488 &2603 &0 &0 &500 &9196 &\\
MinH vs MM: &0 &0 &0 &0 &0 &0 &0 &0 &\\
MaxH vs NRBI: &0 &0 &500 &29846 &0 &0 &500 &30729 &\\
MaxH vs MM: &0 &0 &500 &27243 &0 &0 &500 &21533 &\\
NRBI vs MM: &488 &2603 &0 &0 &500 &9196 &0 &0 &\\ 
\hline
Voss vs MinH: &0 &0 &700 &45658 &0 &0 &700 &54858 &\\
Voss vs MaxH: &6 &6 &654 &2886 &1 &2 &699 &9311 &\\
Voss vs NRBI: &0 &0 &216 &253 &0 &0 &391 &694 &\\
Voss vs MM: &6 &6 &654 &2886 &1 &2 &699 &9311 &\\
MinH vs MaxH: &700 &42778 &0 &0 &700 &45549 &0 &0 &H=7\\
MinH vs NRBI: &1 &1 &672 &3134 &0 &0 &700 &10003 &\\
MinH vs MM: &0 &0 &0 &0 &0 &0 &0 &0 &\\
MaxH vs NRBI: &0 &0 &700 &45911 &0 &0 &700 &55552 &\\
MaxH vs MM: &0 &0 &700 &42778 &0 &0 &700 &45549 &\\
NRBI vs MM: &672 &3134 &1 &1 &700 &10003 &0 &0 &\\ 
\hline
Voss vs MinH: &0 &0 &1000 &66914 &0 &0 &1000 &81809 &\\
Voss vs MaxH: &9 &9 &935 &3952 &0 &0 &1000 &13594 &\\
Voss vs NRBI: &0 &0 &257 &309 &0 &0 &393 &491 &\\
Voss vs MM: &9 &9 &935 &3952 &0 &0 &1000 &13594 &\\
MinH vs MaxH: &1000 &62971 &0 &0 &1000 &68215 &0 &0 &H=10\\
MinH vs NRBI: &0 &0 &959 &4252 &0 &0 &1000 &14085 &\\
MinH vs MM: &0 &0 &0 &0 &0 &0 &0 &0 &\\
MaxH vs NRBI: &0 &0 &1000 &67223 &0 &0 &1000 &82300 &\\
MaxH vs MM: &0 &0 &1000 &62971 &0 &0 &1000 &68215 &\\
NRBI vs MM: &959 &4252 &0 &0 &1000 &14085 &0 &0 &\\
\hline
\end{tabular}
\end{adjustbox}
\end{center}
\caption{Steiner results on c20 and d20 instances with 300 basic vertices}
\end{table}
 For $H=10$, NRBI gets better min cost than Voss in 64.48$\%$ of 1000 vectors. The cost difference of them is 4053.5 in average. In the remaining percent of the vectors the min costs returned by both Voss and NRBI are equal. Voss on no cases is better than NRBI. For this hop,  we also saw that in 97.02$\%$ of 1000 vectors, MM is better than Voss with cost difference 17401.75 and in 1.76$\%$ vectors Voss is better with cost difference 82.12. On the remaining of vectors, both algorithms found same cost.
\begin{table}[H]
\begin{center}
\begin{adjustbox}{width=\textwidth,totalheight=\textheight,keepaspectratio}
\begin{tabular}{cc c c c c c c c c c}\\
\hline 
& &&\multicolumn{2}{c}{MinH} &\multicolumn{2}{c}{MaxH} &\multicolumn{2}{c}{MM} &\multicolumn{2}{c}{NRBI} \\ 
\cline{4-11} 
Inctances&Hop &Voss &MinCost &Imp$\%$ &MinCost &Imp$\%$ &MinCost &Imp$\%$ &MinCost &Imp$\%$ \\ 
\hline
&3& 92.91 &90.24 &2.87 &90.18 &2.93 &90.12 &3 &89.80 &3.34\\ 
$\{c5,d5\}$& 5 &517.63 &508.81 &1.70 &514.58 &0.58 &494.26 &4.51 &492.47 &4.86\\
$\{c10,d10\}$& 7 &889.99 &867.31 &2.54 &870.79 &2.15 &831.67 &6.55 &825.64 &7.23\\
& 10 &1044.39 &972.10 &6.92 &981.38 &6.03 &944.72 &9.54 &928.45 &11.10\\
\hline
& 3 &446.06 &438.25 &1.75 &443.35 &0.6 &415.46 &6.86 &412.39 &7.54\\
$\{c15,d15\}$& 5 &420.25 &367.43 &12.56 &375.19 &10.72 &348.80 &17.00 &345.09 &17.88\\
$\{c20,d20\}$& 7 &378.12 &304.00 &19.60 &302.72 &19.94 &296.90 &21.40 &292.11 &22.70\\
& 10 &361.05 &265.70 &26.40 &356.03 &1.39 &24.08 &33.20 &235.907 &34.60\\ 
\hline
\end{tabular}
\end{adjustbox}
\end{center}
\caption{Performances of proposed algorithms in comparison to Voss}
\end{table}
Now we report optimality of algorithms on two groups of graphs, sparse and dense. The analysis in Table 9 indicates that NRBI greedy algorithm performs quite well and consistent. The column $"Imp"$ indicates the amount of improvement over the minimum objective function of all proposed algorithms to Voss. As the density  of a graph and number of hop increases, its performance improves. The average of improvements of NRBI to Voss on sparse graph instances are presented for all hops and it is in the range of 3.34 to 11.10. 
The highest improvement on these graphs is when the hop is 10. Also, For dense graphs the improvement is between 7.54 to 34.60 when the hop is 10. This amount for MM which is combination of MinHIG and MaxHIG is between 3 to 9.54 on sparse graphs and 6.86 to 33.20 on dense graphs. Table 9 also shows all improvements of MinHIG and MaxHIG algorithms on both sparse and dense instances with hops 3, 5, 7, and 10.

Figure 4 shows an average decrease of cost of all algorithms on all different hops, where one can observe that the average amount of improvement on all hops for MM and NRBI algorithm to Voss.
\begin{figure}[H]
\centering
\includegraphics[scale=0.45]{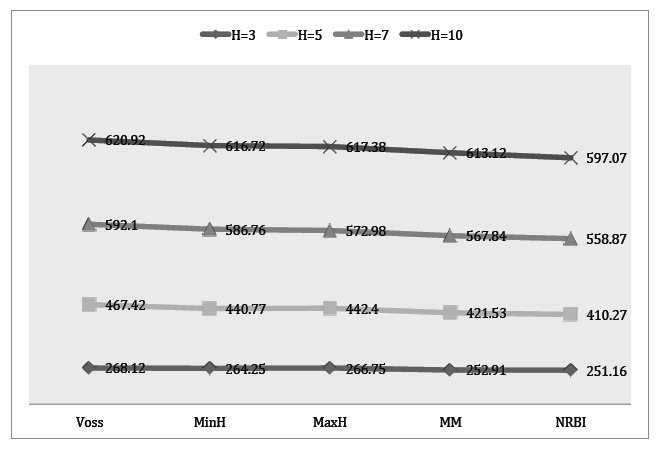}
\caption{Average of Minimum cost found with algorithms on H=3, 5, 7, 10.}
\label{figNNfdsds}
\end{figure}
\section{Conclusion}
\label{}
In this paper we have proposed three greedy algorithms to solve Steiner tree problem with hop constraint, which is a important class of network design. The basic ideas of first two greedy algorithms, MinHIG and MaxHIG, were limited to prim algorithm. They were good on some graphs, but we could not arrive to near optimal solution on all kind of graphs. In fact, the structure of graphs has effect on these algorithms' solutions, since the algorithms are root based. 
Furthermore, we proposed an algorithm ,NRBI, which expand the idea of Kruskal algorithm. The algorithm seemed to be quite robust even when more generalized problems are considered, such as having different hop constraints for all basic vertices. In addition, some comprehensive analysis of all proposed algorithms and the comparison to the Voss's algorithm has been shown that  NRBI algorithm arrived to best solution almost in all cases.  The improvement is significantly 34.60$\%$ in the best case when hop is 10 on dense graphs and in the worst case 3.34$\%$ for hop 3 on sparse graphs on these problems. An interesting future work could be use of heuristic algorithms to try to improve the feasible solutions of the proposed greedy algorithms.



\begin{thebibliography}{}
\bibitem{103}
Balakrishnan A, Altinkemer K. Using a hop-constrained model to generate alternative communication network design. ORSA Journal of Computing 4, 1992; 192-205.
\bibitem{1}
Costa AM, Cardon JF, Laporte G. Models and branch-and-cut algorithms for the Steiner tree problem with revenues budget and hop constraints. Networks 2009; 53(2):141-159.
\bibitem{2}
Costa AM, Cordeau J, Laporte G. Fast heuristics for the Steiner tree problem with revenues, budget and hop constraints. European Journal of Operational Research 2008. 68-78.
\bibitem{3}
Kruskal B. The Shortest spanning subtree of a graph and the Traveling Salesman Problem. In: Proceedings of the American Mathematical Society. 1956. 48–50. 
\bibitem{4}
Duin C, Voss S. Steiner tree heuristics – A survey. In: H. Dyckoff et al. (Hrsg): Operations Research Proceedings, Springer, Berlin u.a., 1993, 485- 496.
\bibitem{5}
Dahl G, Gouveia L, Requejo C. Formulations and methods for the hop-constrained minimum spanning tree problem. In: P. M. Pardalos and M. Resende, editors, Handbook of Optimization in Telecommunications, 2006, 493-515. Springer.
\bibitem{111}
Akgun I, Tansel BC. Degree constrained minimum spanning tree problem: New formulation via Miller-Tucker-Zemlin constraints. Research Report, Bilkent University, Department of Industrial Engineering, Bilkent, Ankara 2009.
\bibitem{105}
Ljubic I, Gollowitzer S. Layered Graph Approaches to the Hop constrained Connected Facility Location Problem, INFORMS Journal on Computing, vol. 25 no. 2013; 2 256-270.
\bibitem{6}
Beasley JE. In., Distributing test problems by electronic mail. Journal of the Operational Research Society, 1990; 1072-1069.
\bibitem{65}
De Boeck J. Fortz B., Extended formulation for hop constrained distribution network configuration problems. European Journal of the Operational Research Society, 2018; 488-502.
\bibitem{7}
Gouveia L. Multicommodity flow models for spanning with hop constraints. European Journal of Operational Research; 1996. 178-190.
\bibitem{8}
Gouveia L. Using variable redefinition for computing lower bounds for minimum spanning and Steiner trees with hop constraints. INFORMS Journal on Computing, 1998; 10(2):180-188.
\bibitem{9}
Gouveia L. Using hop-index models for constrained spanning and Steiner tree models. In: B. Sanso and P. Soriano, editors, Telecommunications network planning; 1999. 21-32.
\bibitem{10}
Gouveia L. Using the Miller-Tucker-Zemlin constraints to formulate a minimal spanning tree problem with hop constraints. Computers Operations Research, 1995; 22(9): 959-970.
\bibitem{102}
Gouveia L, Magnanti TL. Network flows models for designing diameter-constrained minimum spanning and Steiner trees. Networks 41, 2003; 159-173.
\bibitem{11}
Gouveia L, Leitner M, Ljubić I. On the Hop Constrained Steiner Tree Problem with Multiple Root Nodes. Combinatorial Optimization Lecture Notes in Computer Science 2012; 201-212.
\bibitem{110}
Gouveia L, Paissa A, Sharmab D. Modeling and solving the rooted distanceconstrained minimum spanning tree problem. Computers and Operations research Lecture, 2008; 35: 600-613.
\bibitem{101}
Leblance L, Chifflet J, Mahey P. Packet routing in telecommunication networks with path and flow restrictions. INFORMS Journal in Computing 11, 1999; 188-197.
\bibitem{107}
Leblance L, Reddoch R. Reliable link topology/capacity design and routing in backbone telecommunications networks. Working paper, Vanderbilt University, presented at 1st ORSA SIG Conf. on Telecommunications, 1990.
\bibitem{108}
Leblance L, Reddoch R, Chifflet J, Mahey P. Routing in telecommunication networks with flow restrictions. Working paper, Vanderbilt University, 1995.
\bibitem{109}
Maculan N.The Steiner problem in graphs. Annals of Discrete Mathematics, 1987; 31: 185-212.
\bibitem{12}
Prim RC. Shortest connection networks and some generalizations, Bell syst. Techn. J. 1957; 36:1389-1401.
\bibitem{13}
Voss S. Steiner$^\prime$s problem in graphs: Heuristic methods, Discrete Applied Mathematics; 1992 40:45-72.
\bibitem{14}
Voss S. The Steiner tree problem with hop constraint. Annals of Operations Research 1999; 321-345. 
\bibitem{15}
Dokeroglu T., Mengusoglu E.. A self-adaptive and stagnation-aware breakout local search algorithm on the grid for the Steiner tree problem with revenue, budget and hop constraints. Soft Comput. 2018, 22:4133?4151. 
\end{thebibliography}
\end{document}